\documentclass[12pt]{article}
\usepackage[usenames,dvips]{pstricks}
\usepackage{color}
\usepackage{amssymb}
\usepackage{footnote}
\usepackage{longtable}
\usepackage{verbatim}
\usepackage{array,multirow}
\usepackage{titlesec}
\usepackage{physics}
\usepackage{lineno}
\usepackage{url}
\usepackage{placeins}

\usepackage{graphicx}
\usepackage{caption}
\include{symbols}
\def\microns{$\mu$m~}
\def\pd{\ensuremath{\partial }}
\def\LAPPDTM{LAPPD$^{TM}$~}
\newrgbcolor{maroon}{.45 0.1 0.1}
\begin{document}
\pagestyle{plain}

\begin{center}
{\Large\bf Segmented Anodes with Sub-millimeter Spatial Resolution for MCP-Based Photodetectors}
\end{center}

\vskip-0.5in

\begin{center}
Jinseo Park, Fangjian Wu, Evan Angelico, Henry J. Frisch, Eric Spieglan\\
{\it Enrico Fermi Institute, the University of Chicago}\\
\today\\
\end{center}


\begin{abstract}
Micro-channel-plate-based photo-detectors are unique in being capable of covering areas of many square-meters while providing sub-millimeter space resolution, time resolutions of less than 10 picoseconds for charged particles, time resolutions of $\approx$30-50 psec for single photons. Incorporating a capacitively-coupled anode allows for the use of external pickup electrodes optimized for occupancy, rate, and time/space resolution. The signal pickup antenna can be implemented as a printed circuit card with a pattern chosen to match the specific application needs. The electrode elements are typically either a 2-dimensional array of pads for high-occupancy/high-rate applications, or a 1-dimensional array of strips for low-occupancy/low-rate, and a lower channel count. Here we present pad patterns that enhance charge-sharing between pads to significantly lower the required channel count/area while maintaining spatial resolutions of $\approx 100$ to 200 microns for charged particles and $\approx 400$ microns to 1 mm for single photons. Patterns that use multiple signal layers in the capacitively-coupled printed circuit signal pickup board can lower the channel count even further, moving the scaling behavior in the number of pads versus total area from quadratic to linear.
\end{abstract}

%

\newpage
\section{Introduction}
\label{introduction}

The precise detection of photons and charged particles over large areas with sub-mm space resolution and time resolutions measured in tens of picoseconds (psec) ~\cite{Ohshima_2006,Vavra_TestBeam_2009,Anatoly_TestBeam_2010,history_paper} enables 3-dimensional imaging by time-of-flight: the arrival space-time coordinates of detected photons may constrain the image by their reconstructed transit times~\cite{OTPC_paper,Oberla_thesis}.

 Monolithic, unsegmented anodes that are capacitively coupled through a dielectric anode substrate are used to allow signal-pickup electrodes to be placed external to the detector vacuum package, enabling batch production of one photodetector design for multiple applications~\cite{RoentDek_strip_patent, RoentDek,  Jagutzki_1999,Jagutzki_2002,Photek_2005,Photek_2007,Jagutzki_2013,Torch_PhD_Lausanne,Torch_MultiAnode_PMT}.  The fast rise times and higher gains inherent in ALD-coated MCP-PMT signals~\cite{Arradiance,Jeff_ALD_ECS_2013,Anil_ALD_ECS_2014, ALD_patent,Slade_SEY_NIM,Minot_Incom_2020} also allow the use of a metal internal anode, with the resistance of the thin-metal layer being high enough to form a high-pass RC filter for signals transmitted through the vacuum package wall~\cite{InsideOut_paper}.

 The external signal-pickup antenna can be easily implemented as a printed circuit card with a customized pattern of signal pickup electrodes. The electrode elements are typically implemented as either a 2-dimensional array of pads for high-rate/occupancy applications, or, if low-rate/occupancy, a 1-dimensional array of strips with a significantly lower channel count~\cite{Tang_Naxos,timing_paper}.  Here we consider the 2-dimensional case of pad patterns with enhanced charge sharing to lower the required channel count per area without significant loss in spatial or temporal resolution. 
 
Because the fundamental spatial scale of
an MCP is the pore lattice spacing, typically 10-25 microns in large-area detectors~\cite{Minot_Incom_2020}, the spatial resolution is determined by the segmentation of the anode plane that measures signals from the charge cloud generated by the MCP pores. However a segmentation with scale size comparable to the time resolution~\cite{units_c_equal_1} would result in millimeter-sized pads, necessitating a channel
count on the order of $\ge10^5$ per m$^2$. The signal amplitude also decreases as the pad size becomes small.

\begin{figure}[th]
\centering
\includegraphics[width=0.85\textwidth]{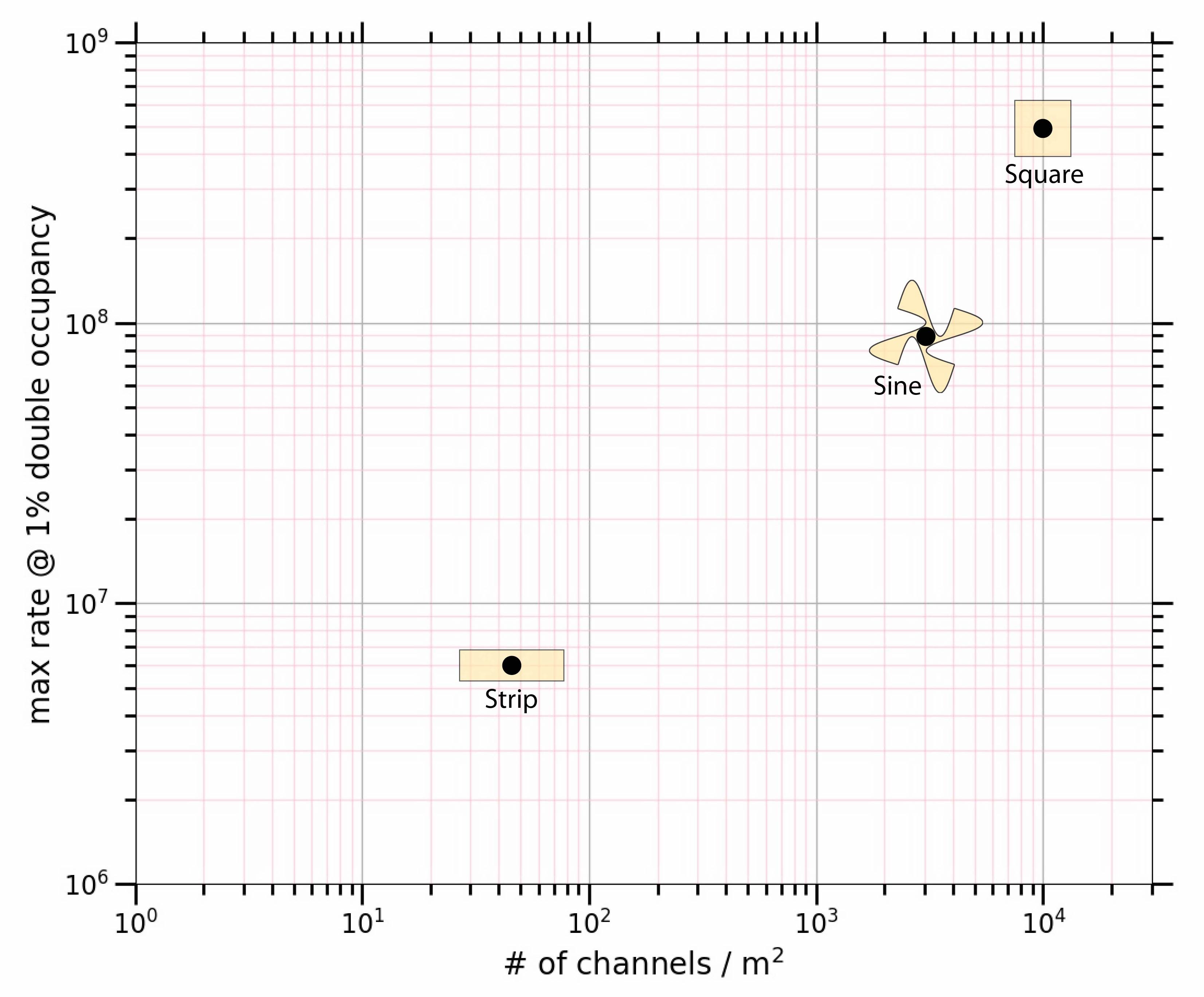}
\caption{The maximum rate of hits-per-channel for an occupancy less than 1\%, versus the number of signal pickup channels per m$^2$ for square pads, strips, and sinusoidal pads with enhanced sharing. The maximum rate $R$ for 1\% occupancy is determined by $R\Delta T = 0.01$, where $\Delta T$ is the time a signal takes to propagate across a pad or strip in the pattern.}
\label{fig:occupancy}
\end{figure}

Figure \ref{fig:occupancy} is a rough representation of the maximum rate of hits a channel could handle versus the number of channels per square meter for square pads, strips, and pads with enhanced sharing. The maximum rate is defined so that the double-hit occupancy per channel is 1\% at it. Here occupancy refers to the number of channels with signals in the resolving time window; it may be low or high in both low-rate and high-rate applications~\cite{occupancy_vs_rate}. For large-area, low-occupancy applications, the single-ended strip patterns are most efficient while still providing sub-mm resolution~\cite{OTPC_paper}.

\section{The Image of the Charge Distribution from Signals in MCP-based Photomultipliers}
\label{charge_image}

The result of amplification of photoelectrons in ALD-functionalized micro-channel plates is a cloud of $\ge 10^7$ electrons exiting the MCP pores. Typically  $\approx$~6 MCP pores are involved in the amplification process when two-MCPs are stacked in the Chevron configuration \cite{wiza}. The electrons have a spread in momentum and angle that causes the avalanche to diverge as it travels to the anode. As the cloud of charge approaches the metal anode, image currents are induced in the anode in response to  the electro-magnetic field lines terminating at the surface.

 Signals induced on a segmented anode can be analyzed  to constrain the position of the incident particle. Consider the case of a square-pad array: if the charge cloud image size is circular (i.e. normal incidence) and the pad size has been set equal to the image diameter, then an event with only one pad containing signal corresponds to a  particle detection at the center of that pad in the photocathode plane. Similarly, if two pads equally share the signal, the cloud image is centered on the boundary in that dimension. The combination of signal sharing and knowledge of the image shape can lead to a measurement of the position of the image at much higher resolution than the pad size.

The spatial dimensions of the charge cloud image are largely set by the spatial dimensions of the charge avalanche~\cite{saito_cloud_size}. Among the factors playing a role in determining the transverse dimensions of the signals induced in the anode are: 1) the length of the gap and voltage between the entry and exit MCP; 2) the length of the gap and voltage between the exit MCP and the anode; 3) the end-spoiling of the MCP pores; 4) the pore bias angle; and 5) the geometry of the pickup pattern. In the case of a capacitively coupled anode, the resistance of the internal anode layer as well as its distance from the pickup plane will also affect the size of the image.

Given a specific detector and signal source, the size and shape of the induced image charge distribution can be measured as input for optimization. We present the simulation results in terms of a scaling parameter $L$, the ratio of pad size to the (detector-dependent) charge image diameter.

We consider signals induced by a single photoelectron, and by a charged particle producing Cherenkov light in the window~\cite{Credo_Rome_2004}. 

\subsection{The image from a single photoelectron}

 Measurements of the image radius from a single photoelectron vary based on the design of the detector as well as the measurement configuration. Signal distributions are typically fit to a Gaussian in either voltage or time-integrated voltage with the standard deviation representing the transverse size. These image sizes vary from 0.5 - 5 mm radius in the case of single electrons \cite{saito_cloud_size, ossy_cloud_size}. The radii in these measurements are most strongly dependent on the separation distances of the active layers in the MCP-PMT and the applied voltages that accelerate the electron cloud \cite{saito_cloud_size}.

\subsection{The image from Cherenkov photons generated by a charged particle}
\label{charged_particle_image}
\begin{figure}[!ht]
\centering
\includegraphics[width=0.65\textwidth]{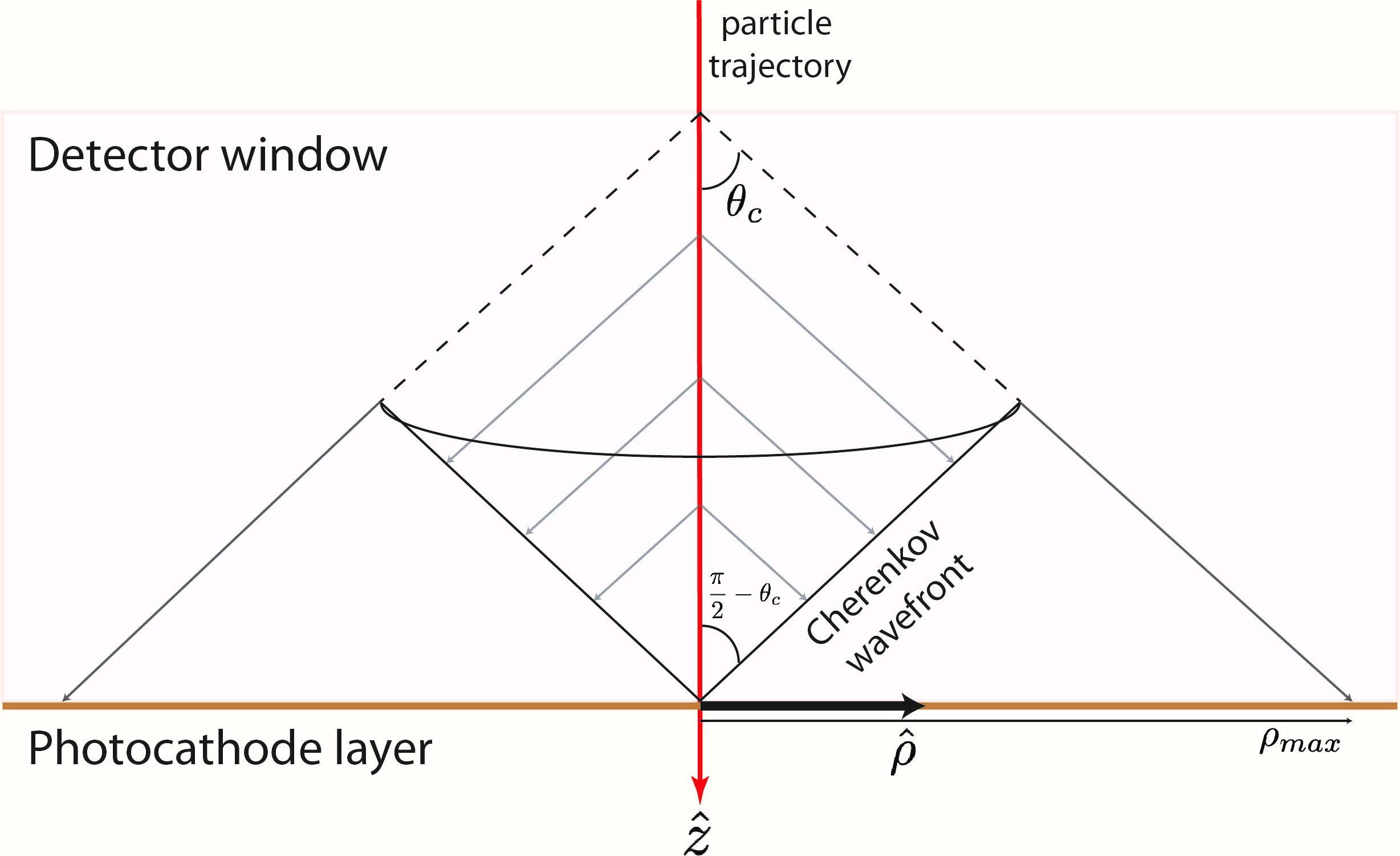}
\caption{The generation of Cherenkov light by a relativistic charged particle traversing the entrance window/radiator at normal incidence. Photons arrive at the photocathode with a maximum radius related to the Cherenkov angle and the window thickness, typically on the order of 0.5 - 1 cm. The pattern of photons is transferred by proximity focusing to the pores in the top MCP, and then is amplified by the MCP plates to produce a charge cloud that forms a circular image at the anode.}
\label{fig:evan_cherenkov_diagram}
\end{figure}

\begin{figure}[!ht]
\centering
\includegraphics[width=0.85\textwidth]{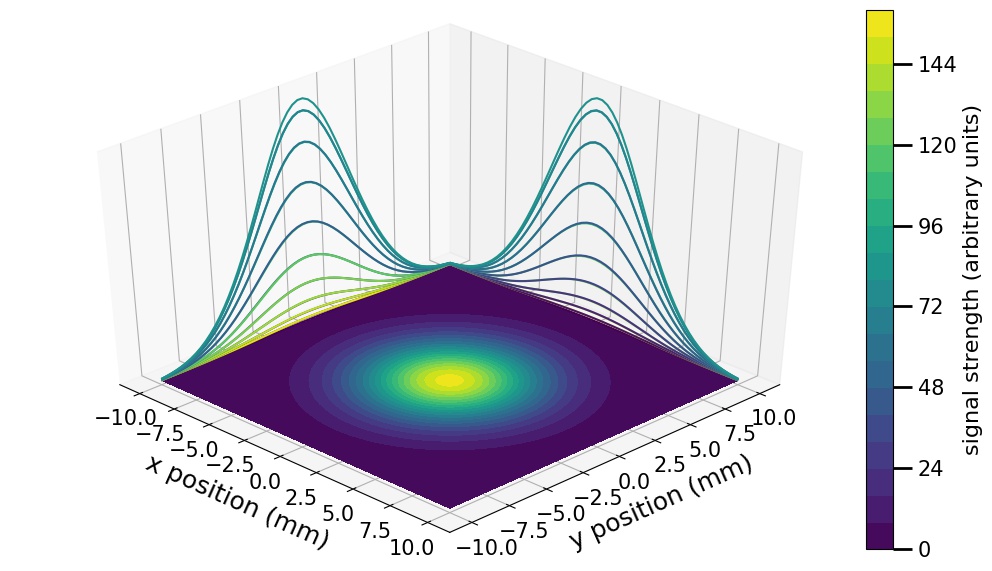}
\caption{The average image of the charge cloud generated by a normally-incident charged particle. The single photoelectron image is modeled as a Gaussian of 5 mm standard deviation, and the Cherenkov photons are generated over 5 mm of radiator. The Cherenkov angle is fixed at 48$^{\circ}$ and the number of photoelectrons per centimeter of glass is 66. These parameters are motivated by the characteristics of LAPPDs used in Reference \cite{Angelico_thesis}. The average is over 300 events.}
\label{fig:image-shape-model}
\end{figure}

Charged particles may be detected by LAPPDs using Cherenkov light produced in the front window, as shown in Figure \ref{fig:evan_cherenkov_diagram}. The arrival of Cherenkov photons is a good proxy for the arrival-time of charged particles as Cherenkov emission preserves timing at the sub-picosecond level \cite{jelley}. After traveling through the glass window at the Cherenkov-emission angle, photons originating on the particle trajectory are converted to photoelectrons at the photocathode. The photoelectrons are then proximity focused to the pores of the MCP where they interact to initiate a shower. Here we assume normal incidence of the charged particle; off-angle incidence will produce measurable, and hence exploitable, effects depending on the anode pattern. Optimizing for angular resolution is beyond the present scope of this paper.

Cherenkov photons are emitted at an angle $\theta_c$ such that $\cos \theta_c = 1/\beta n(\lambda)$ \cite{jelley}, where the index of refraction $n(\lambda)$ is wavelength-dependent. Charged particles at normal incidence produce a circular spot centered on the transverse position of the particle with a maximum radius $T \tan\theta_{c}$. where $T$ is the thickness of the window or radiator. For Schott B33 glass \cite{B33, Schott}, a typical material for photo-detector windows, the Cherenkov angle ranges from $\theta_c =$ 48.3$^{\circ}$ at 300 nm to 47.1$^{\circ}$ at 700 nm. The resulting Cherenkov photon-spot radius is roughly equal to the thickness of the radiator.

Using the typical performance of bi-alkali photocathodes and the transmission of B33 glass windows, the number of photoelectrons is $\approx$~66 per centimeter of radiator; for fused silica, the number of photoelectrons increases to $\approx$~200 per centimeter due to enhanced ultra-violet transmission  \cite{Angelico_thesis}.

For an internal (as opposed to capacitively coupled) anode consisting of a 1-dimensional array of strip-line conductors with 5.1 mm width and 6.9 mm pitch, the resulting transverse size of the charged-particle signal is $\approx 10 \pm 2.5$ mm FWHM\footnote{This measurement was made with an LAPPD with two 1.3 mm MCPs separated by approximately 2 mm and a distance of about 6 mm between the exit MCP and anode. The MCPs are each biased at 900 V, photocathode at 20 V, and 200 V across each of the gaps.}~\cite{Angelico_thesis}.

A simulation of photoelectron positions produced by a charged particle shows that if the signal from a single photoelectron is Gaussian, the signal from an impinging charged particle will be roughly Gaussian with the exact shape depending on the thicknesses of elements in the detector. Figure \ref{fig:image-shape-model} is an averaged image of the charge cloud for photoelectrons generated by Cherenkov light emitted from a normally-incident charged particle in a 5 mm thick radiator. The shape is dependent on the model for the single photoelectron image, which is taken to be a Gaussian of 5 mm standard deviation.

A Gaussian is used to model the image of the charge cloud in the simulations presented in Section \ref{results}. The size of the image is compared with the size of the individual anode pads by scaling $L$, which is the ratio of the pad-side length and 4$\sigma$ of the Gaussian image.

\section{Sub-Millimeter Position Reconstruction Using Patterned Anodes}
\label{reconstruction}

Capacitive coupling of the anode plane enables the use of printed circuit boards as the external signal-pickup component~\cite{InsideOut_paper}. These are easily printed in complex patterns. Because they are external to and electrically isolated from the photomultiplier,
they can be optimized for specific applications, occupancies, time/space resolutions, and rates. Printed circuit boards are inexpensive and widely available with fast turnaround for rapid optimization.

Spatial resolutions of $\approx 300$ \microns  have been obtained using a signal-pickup board with 12.7 mm-square pads for signal-source locations for which there is charge sharing between neighboring pads~\cite{InsideOut_paper}. However when the image of the charge cloud is fully contained within one pad, the resolution becomes substantially worse depending on the ratio of the pad size to the diameter of the image.

Large-area high-rate applications such as high-energy particle colliders and some medical applications are natural candidates for pad-based signal pickup segmentation. However the number of channels grows quadratically with inverse pad size.  Here we discuss signal pickups with patterns that enhance charge sharing between pads, enabling sub-mm resolutions for pad sizes with characteristic lengths larger than the charge pattern, reducing the channel count quadratically in the ratio of pad size to charge image size.

\subsection{Calculation of spatial resolution}

The response and concomitant spatial resolution for different signal pickup pad implementations are calculated by identifying the spatial resolution with the 2D gradient of the signal distribution over the entire pattern. At many discrete points on a pad pattern, the signal collected by each pad is calculated using the overlap of the image distribution, modeled as a bi-variate Gaussian, with each pad. The collection of all simulated positions forms a look-up table. From the look-up table, a 2D signal gradient is calculated as a function of the charge-cloud position. The derivation of this function for the discrete simulation used here is described in Appendix A.

\subsubsection{The effect of electronic noise on reconstruction using pad sharing}
\label{electronic_noise}

The determination of the image position using the relative sharing between pads has inherent uncertainties due to electronic noise as well as possible fluctuations and non-uniformities in the MCP amplification stages. These sources of noise increase the degree of signal-sharing required for a desired spatial resolution.

Voltage noise of modern-day, fast-sampling waveform digitizers used to measure MCP-PMT signals is on the order of 0.5 - 1 mV. For example, the PSEC4 digitizing ASIC for digitizing LAPPDs has an RMS voltage noise of about 700 $\mu$V \cite{Oberla_thesis, PSEC4_paper}.

Signal amplitudes from single photoelectrons detected by MCP-PMTs can range from 5 - 40 mV depending on the gain of the detector, the capacitance and impedance properties of the anode, the size and shape of the anode pads, and the size and shape of the charge-cloud image. In addition, if the anode is capacitively coupled, there may be a reduction in amplitude due to attenuation through the coupling or due to the spreading of the signal as it propagates to the pickup pattern \cite{InsideOut_paper}.

For an LAPPD with an internal strip-line anode, the noise of the PSEC4 electronics increases from 700 $\mu$V to about 1.5 mV due to the antenna-like properties of the cables and strip lines \cite{Angelico_thesis}. At a gain of 3 $\times$ 10$^{7}$, an LAPPD with the same strip-line anode pattern measured a single photoelectron pulse-amplitude distribution with a peak at 60 mV \cite{timing_paper}.

Because the noise-to-signal ratio varies depending on detector settings and anode configuration, we choose a conservative ratio of 7.5\% which takes into account 1.5 mV noise and a detector with a (modest) gain of 10$^{7}$ gain, resulting in 20 mV single photoelectron amplitude. This is the noise value $\sigma_C$ used in the simulation of single photoelectrons in the results to follow, which enter in the resolution calculation detailed in Appendix A.

A charged particle passing through the window will typically produce many tens of photoelectrons, increasing the signal amplitude to greater than 100 mV. At the limit of high signal amplitude, the limiting factor in reconstructing the position of the charged particle using the method of sharing may be the modeling of the transverse shape and distribution of the image. In the results to follow, resolution functions corresponding to a 1\% noise-to-signal ratio are plotted alongside the single photoelectron case.

The resulting resolution functions scale linearly with the noise-to-signal ratio. The noise levels used here, though based on observations from experimental setups, are somewhat arbitrary due to the dependence on detector settings. The reader may scale the resulting resolutions to correspond to their particular experimental setup.

\section{Anode Pattern Simulation Results}
\label{results}

We apply the algorithm of Appendix A to two patterns: a regular pattern of square pads as the baseline, and a pattern of sinusoidal pads, which was expected to have reduced channel count and more uniform spatial resolution. For a given pattern the spatial resolutions are calculated as a function of the size of the pads via a scale factor, $L$, defined as the ratio of the pad characteristic size (e.g. the side of a square pad) to the diameter of the image of the charge cloud. As an example, $L=2$ corresponds to half as many pads per unit length compared to $L = 1$, and one quarter as many per unit area. The diameter of the image, modeled as a Gaussian, is defined as the size at $4\sigma$ of the Gaussian. The value of $\sigma$ is set to $4.2$mm which corresponds to the $10$mm FWHM described in Section~\ref{charged_particle_image}. The size of the pad is scaled relative to this Gaussian diameter to yield $L$.

%
\begin{figure}[!ht]
\centering
\includegraphics[width=0.4\textwidth]{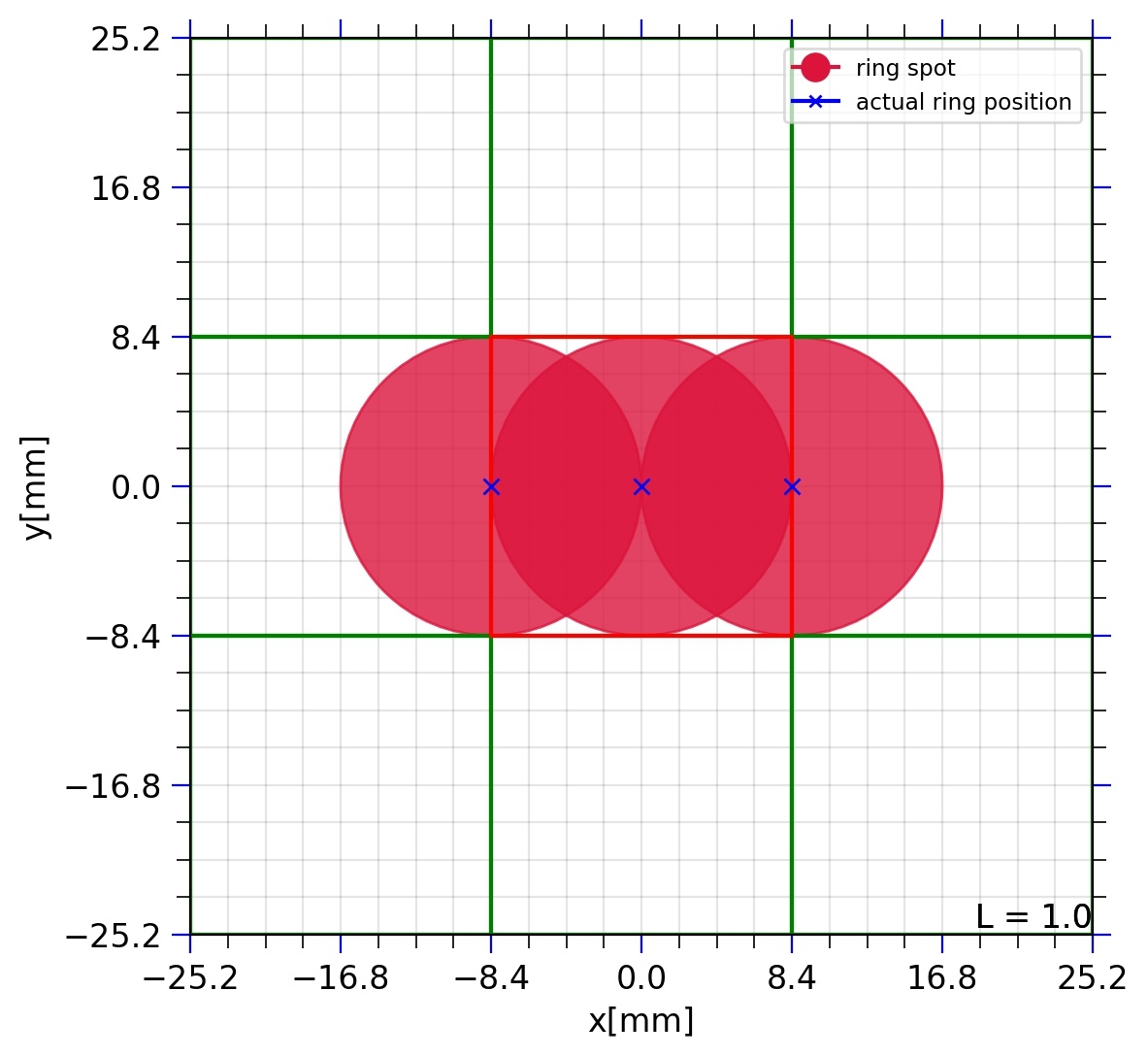}
\includegraphics[width=0.4\textwidth]{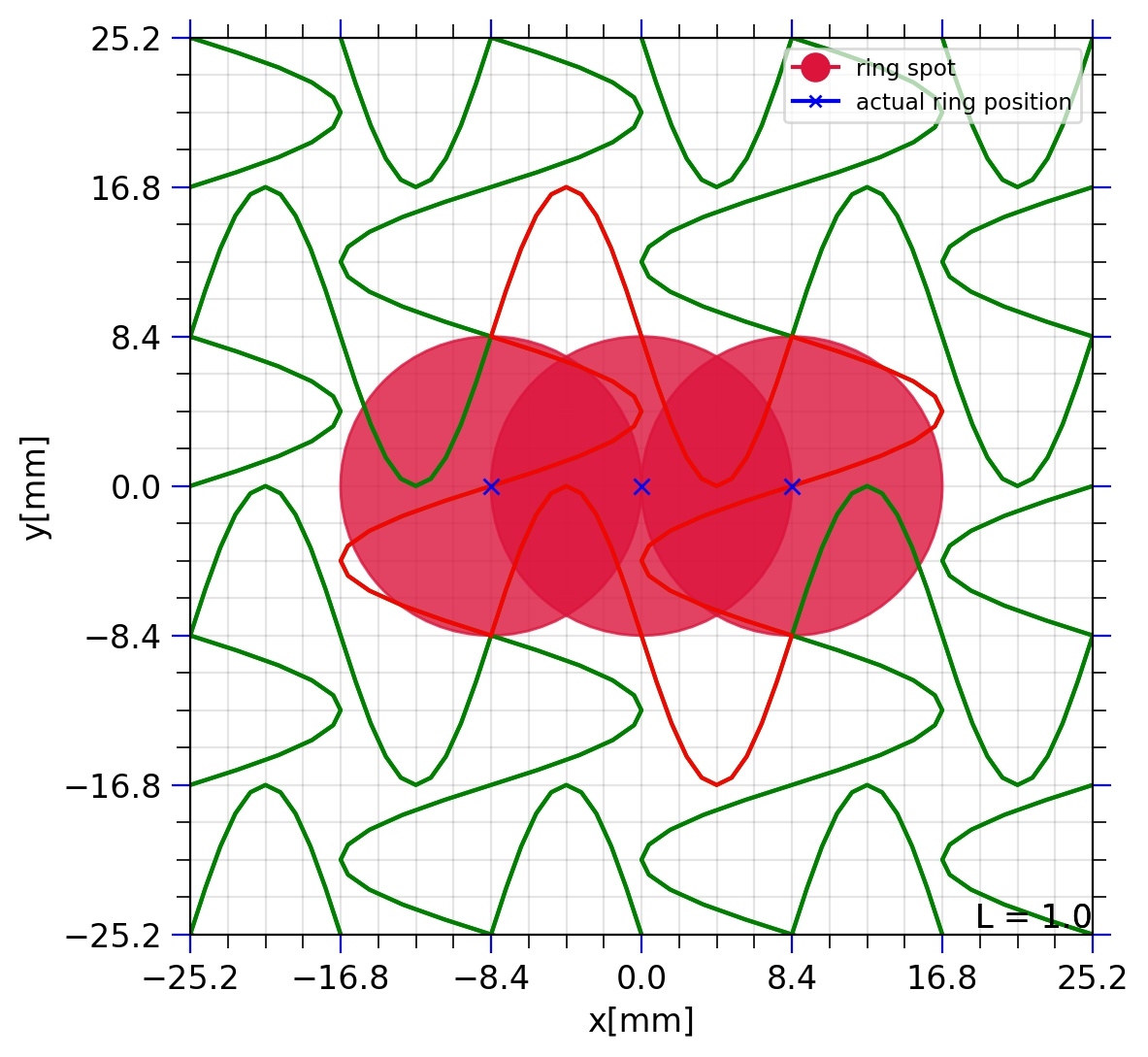}
\caption{Left: Nine cells of a square-pad pattern with scale factor $L = 1.0$. The charge image is shown in three positions: centered on a pad, and on the left and right boundaries of the central pad (outlined in red). Right: The sinusoidal-pad pattern with $L = 1.0$ and $H = 0.5$. The central pad in both patterns is outlined in red for clarity.}
\label{fig:square_sinusoidal_pads}
\end{figure}

\subsection{Square and sinusoidal pads}
A 3 $\times$ 3 portion of a square-pad pattern, like that used in the measurements of Ref.~\cite{InsideOut_paper}, is shown in the left-hand panel of Figure~\ref{fig:square_sinusoidal_pads}. The red circle represents a uniform and circular version of the charge image produced by a single photoelectron or charged particle. The radius of the boundary of the plotted circles represents 2$\sigma$ of the Gaussian image distribution. The ratio of the pad side to the diameter of the charge image, the scale factor $L$, is $1.0$.

One guiding principle in increasing the degree of signal sharing between pads is to distort the boundaries of the square pads such that the image can never fit entirely within one pad. An example is the sinusoidal-pad pattern shown on the right-hand panel of Figure~\ref{fig:square_sinusoidal_pads}. This pattern has the same scale factor but introduces an additional parameter: the amplitude of the sine-wave distortion. Normalized to the pad characteristic size, the amplitude shown in the figure is 0.5.


\subsection{Response of neighboring cells versus incident position}
\label{neighbors}
\begin{figure}[bh]
\centering
\includegraphics[width=0.45\textwidth]{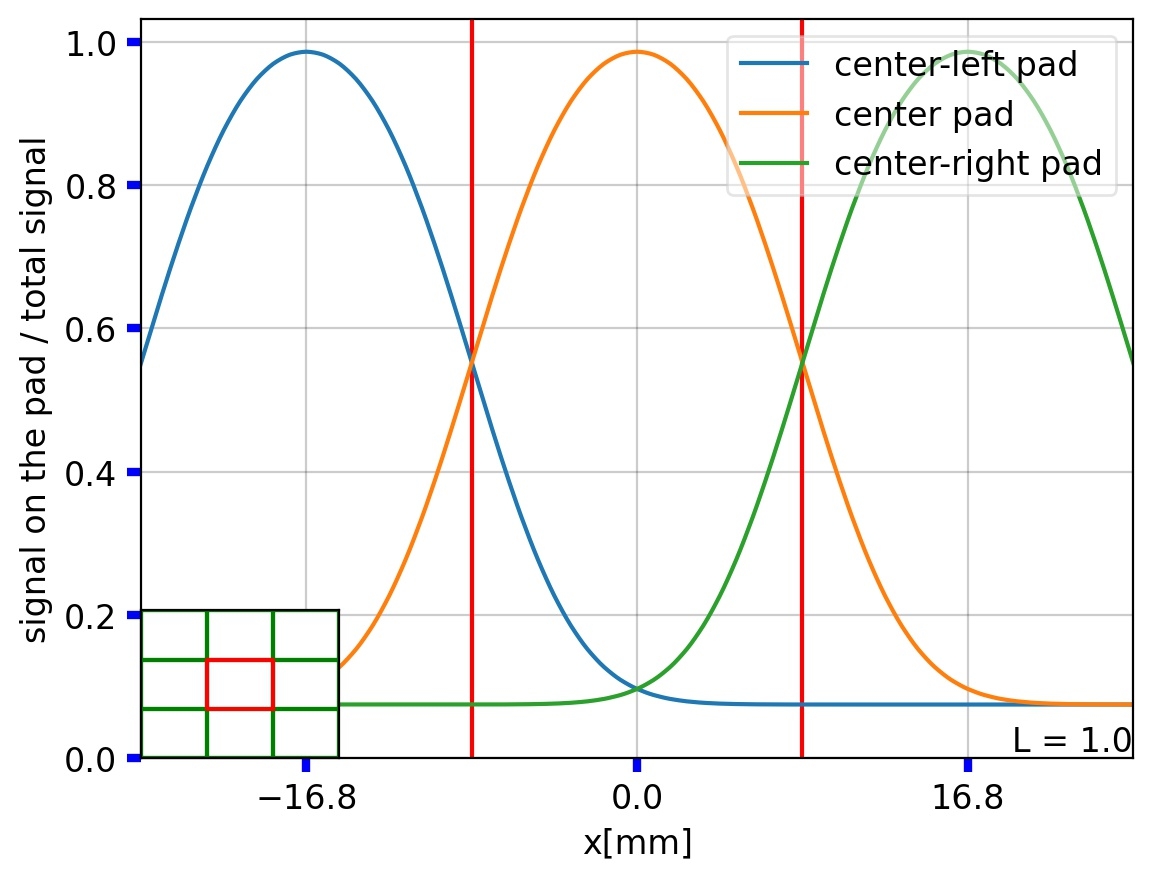}
\includegraphics[width=0.45\textwidth]{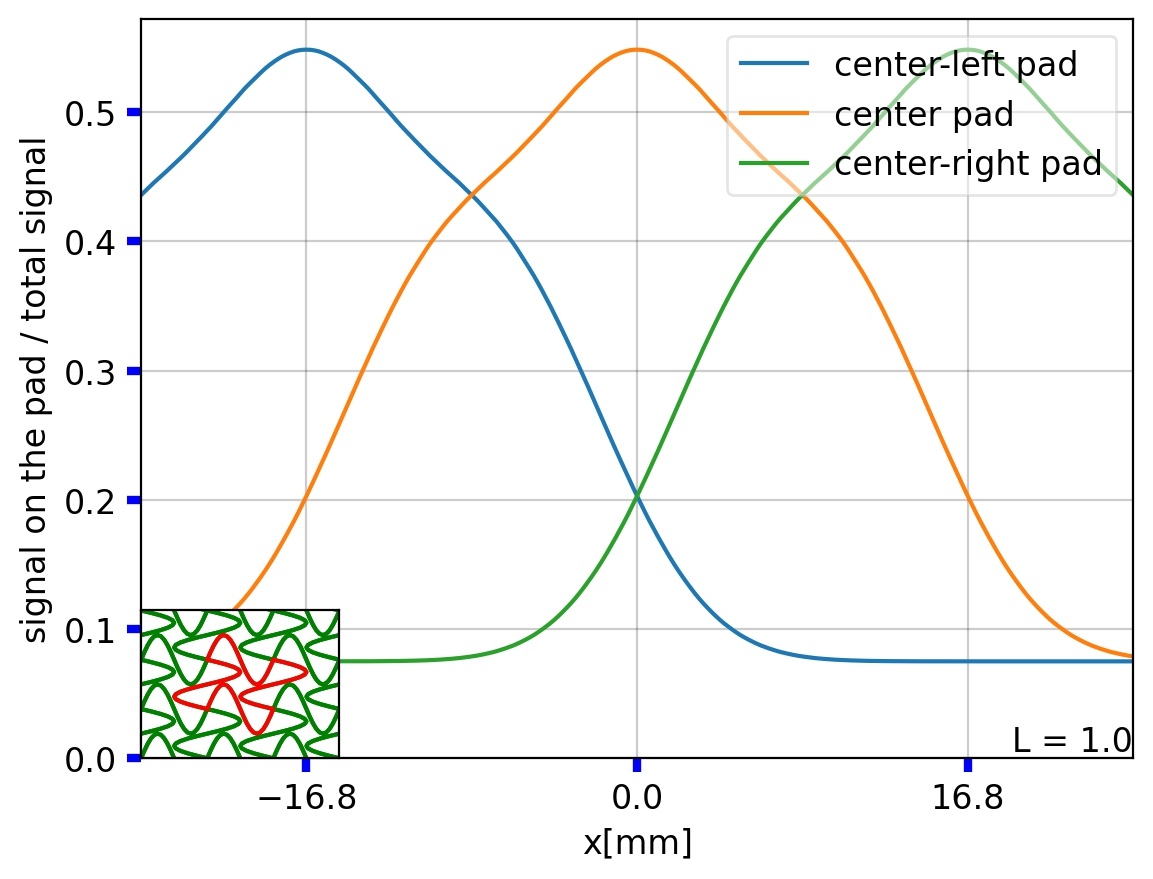}
\caption{ The signal amplitude of the center row of pads as the horizontal position of the image is scanned over the pads. Left: Square pad pattern with $L=1.0$. Right: Sinusoidal pad pattern with $L=1.0$.}
\label{fig:fractional_charge_response}
\end{figure}

%
%
The fraction of the total signal measured on each of two neighboring cells is shown in Figure~\ref{fig:fractional_charge_response} as the image position is scanned across the line $y = 0$. The square pattern traces out the overlap of the image with the equal-size pad, not going to zero or infinity due to the
presence of signal from digitizing-electronics noise with magnitude equal to 7.5\% of the total signal. The right-hand plot shows that the sinusoidal pattern produces a larger degree of signal sharing.

%
%

\begin{figure}[!ht]
\centering
\includegraphics[width=0.4\textwidth]{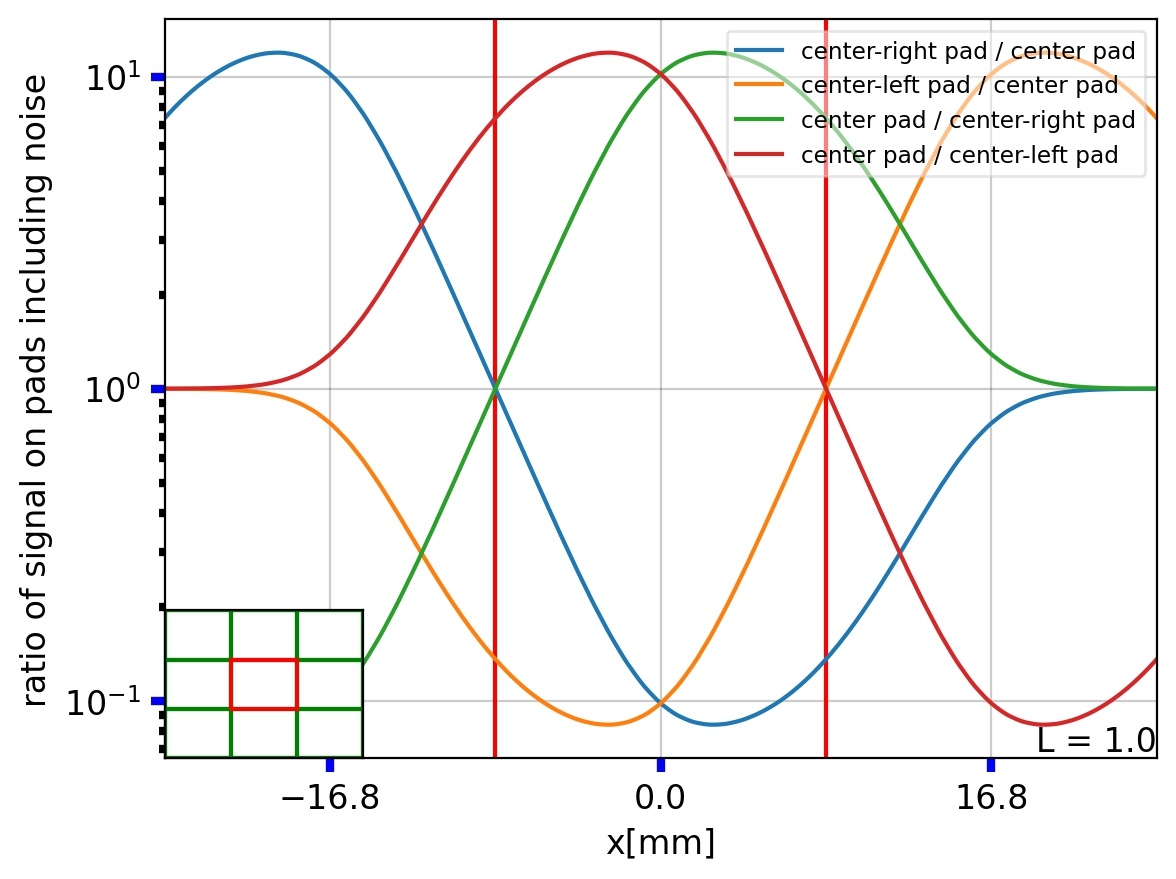}
\includegraphics[width=0.4\textwidth]{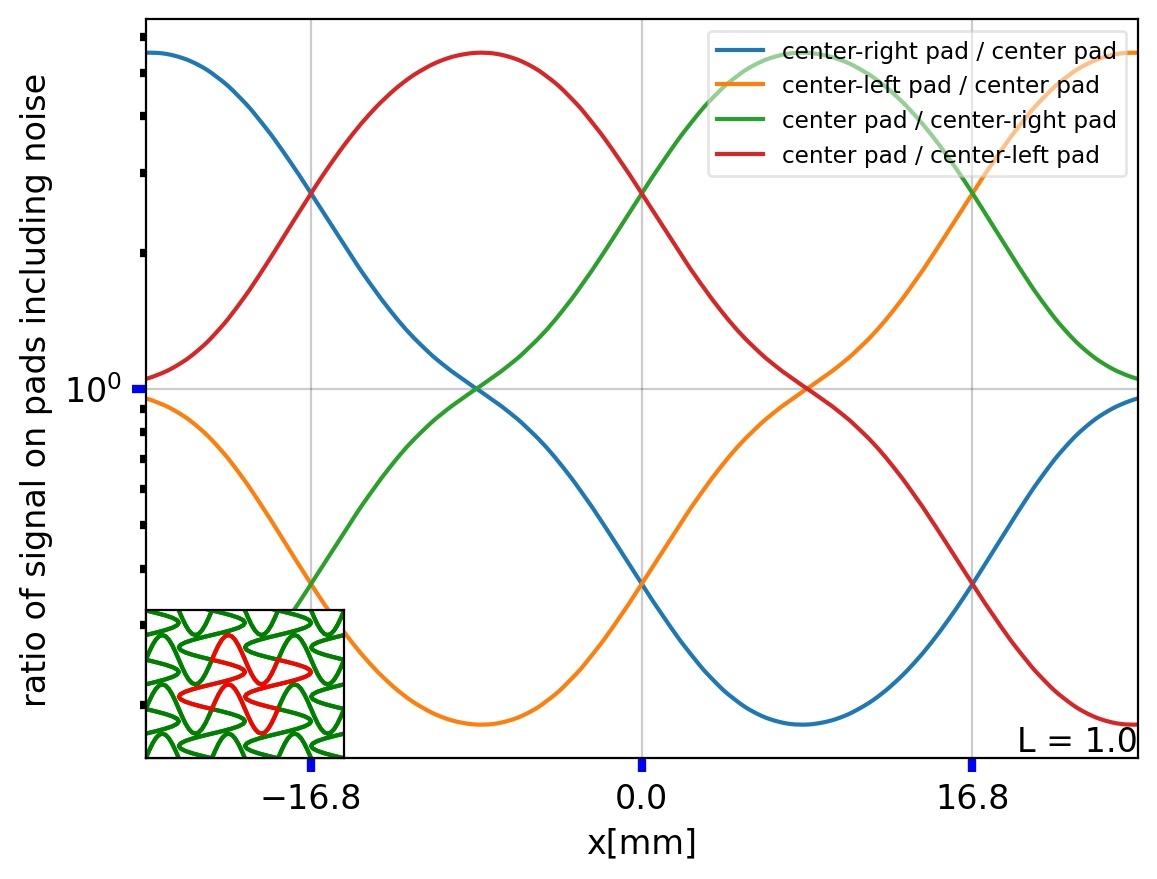}
\caption{The signal ratio of neighboring pads as the horizontal position of the image is scanned across the line $y=0$. Left: Square pad pattern with scale factor $L=1.0$. Right: Sinusoidal pad pattern with $L=1.0$.}
\label{fig:ratio_charge_response}
\end{figure}

%
%
%
The ratio of the charge deposited on each of two neighboring cells as the image position is scanned across the line $y = 0$ is shown in Figure~\ref{fig:ratio_charge_response} for both the square (Left) and sinusoidal (Right) patterns.
%
%
%
%
%

\subsection{Position resolution versus incident particle position}
\label{resolution}

The spatial resolution may be calculated as a function of position on the entire array of pads using the mathematics for the gradient outlined in Appendix A. The resulting spatial resolutions of the square and sinusoidal patterns are presented in this section for a noise-to-signal ratio of 7.5\% (single photoelectrons) and 1\% (charged particles).

All spatial resolutions are reported in microns, but are directly proportional to the noise-to-signal ratio. For example, a resolution of 100 microns shown here in the case of 1\% noise would represent a resolution of 1000 microns in the case of 10\% noise.

\subsubsection{Resolution functions in 2D for fixed L}
\begin{figure}[!ht]
\centering
\includegraphics[width=0.4\textwidth]{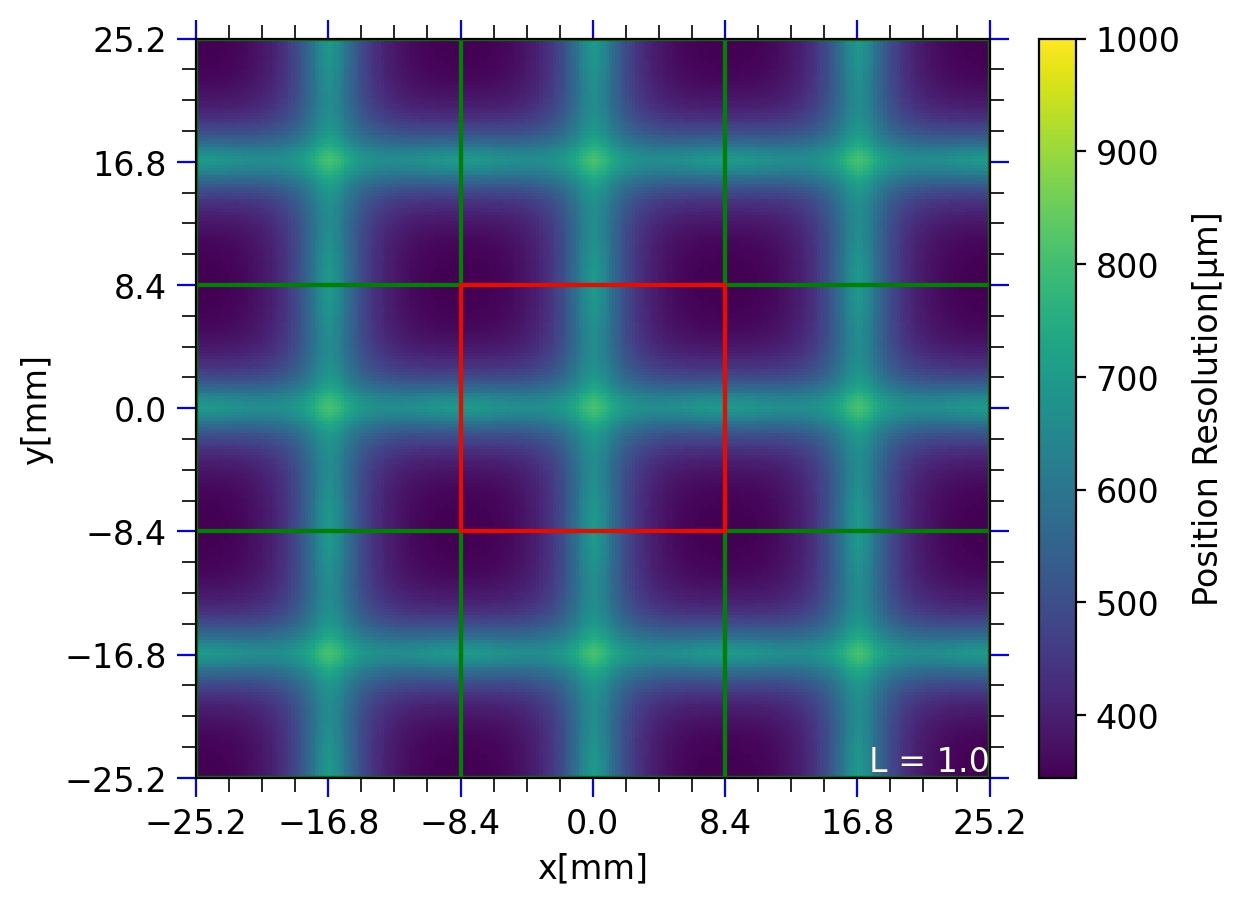}
\includegraphics[width=0.4\textwidth]{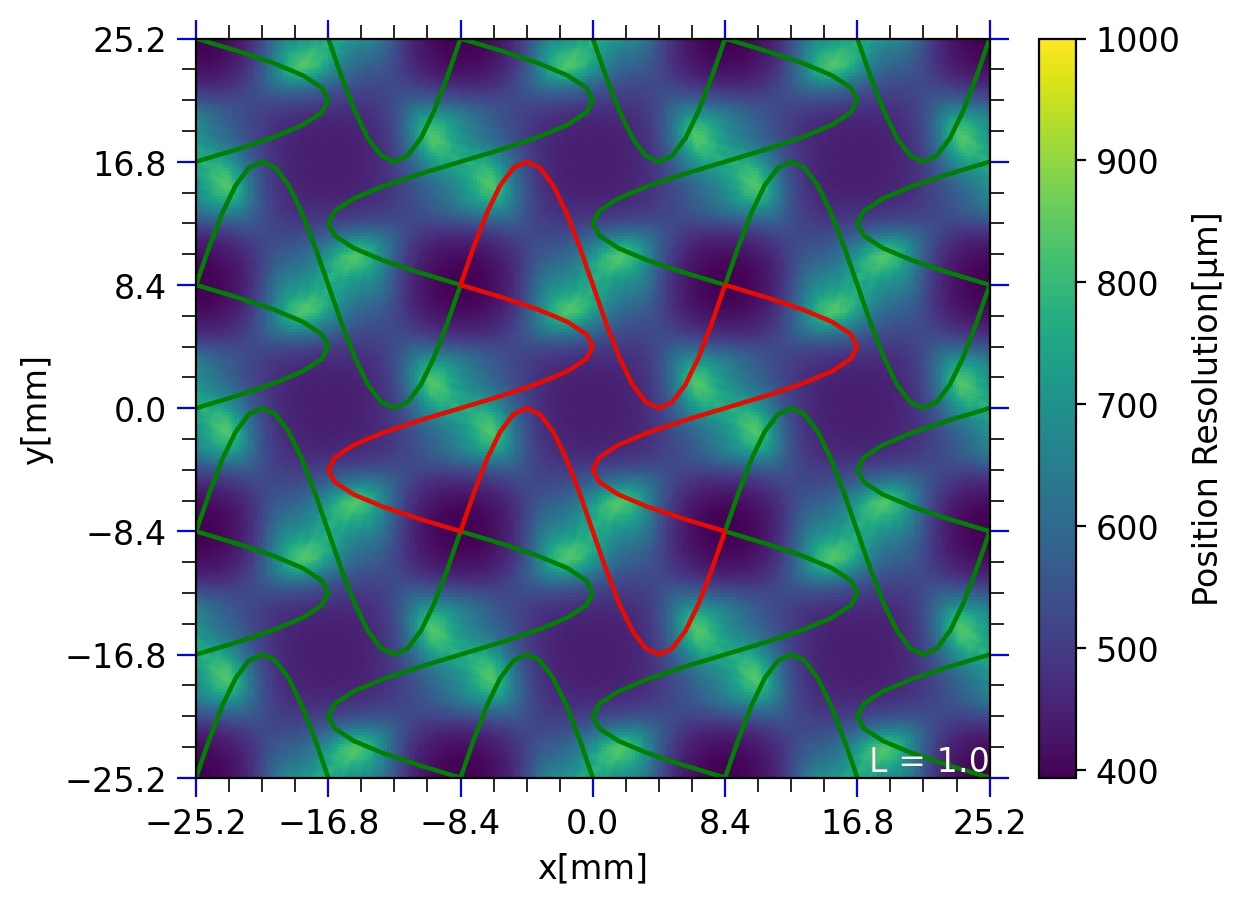}
\caption{The position resolution as a function of incident position for a noise-to-signal ratio of 7.5\%.  Left: Square pad pattern with $L=1.0$. Right: Sinusoidal pad pattern with $L=1.0$.}
\label{fig:2D_resolution}
\end{figure}
The 2D resolution function for the square and sinusoidal patterns is shown in Figure~\ref{fig:2D_resolution}. This resolution is calculated using the signal-gradient method of Appendix A with 7.5\% noise $\sigma_C$. Locations where the resolution peaks represent locations where the local gradient of the signal sharing is smallest. These locations may be used to inform further optimization of the patterns. The best resolution for the square pattern is $\approx 550$ microns and the resolution spikes to $\approx 770$ microns when the image is centered on the pad.

\begin{figure}[!ht]
\centering
\includegraphics[width=0.4\textwidth]{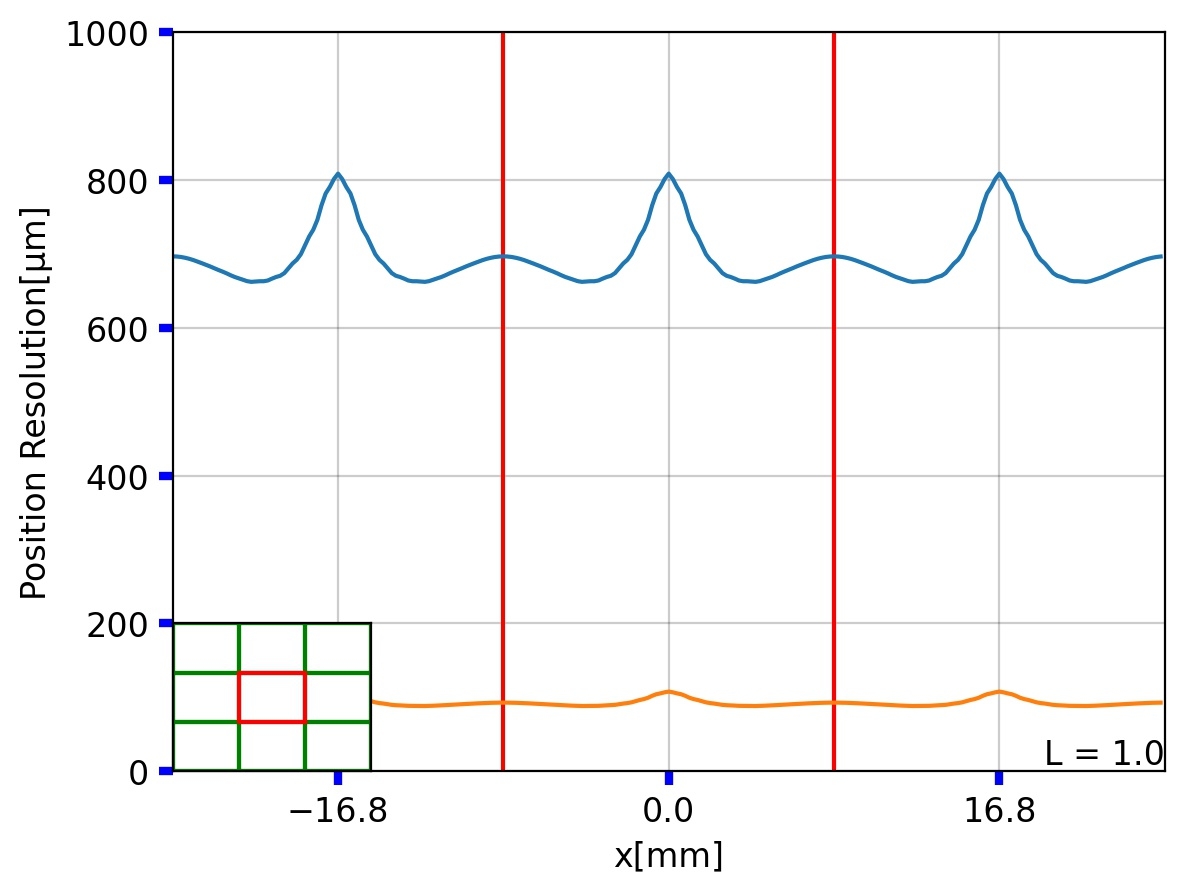}
\includegraphics[width=0.4\textwidth]{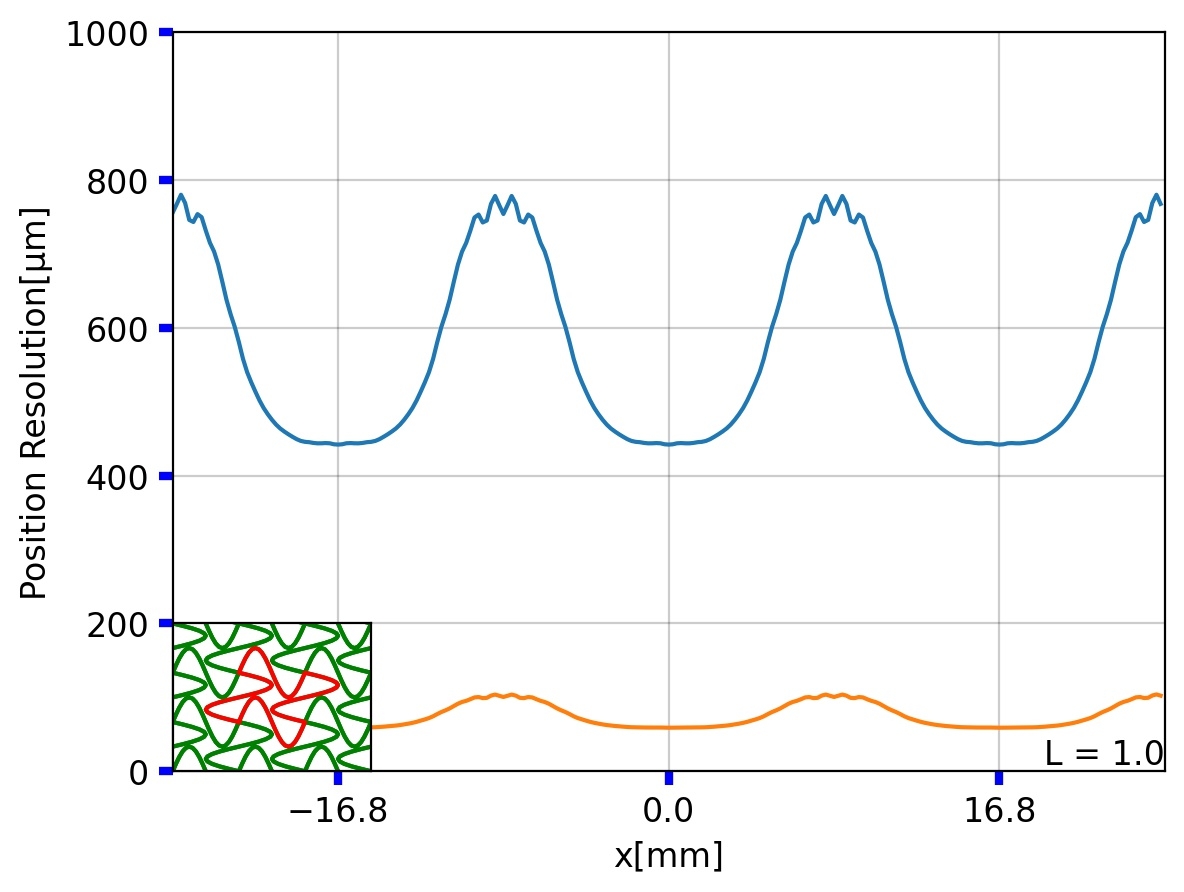}
\caption{A slice of the position resolution functional at $y = 0$. The blue lines represent a noise level of 7.5\% and the orange lines represent a noise level of 1\%. Left: Square pad pattern with $L=1.0$. Right: Sinusoidal pad pattern with $L=1.0$. The ripples on the peaks of the sinusoidal pattern's function are generated by the discretization of the anode, look-up table, and signal shape function.}
\label{fig:1D_resolution}
\end{figure}

A one dimensional slice of the 2D resolution function along the line $y = 0$ is shown in Figure~\ref{fig:1D_resolution}. In the following section, the 10$^{\text{th}}$ and 90$^{\text{th}}$ percentile, as well as the median, of the full 2D resolution functions are reported as a metric of pattern performance.
%

\subsubsection{Resolution as a function of the pad size}
\label{scale}

\begin{figure}[!ht]
\centering
\includegraphics[width=0.4\textwidth]{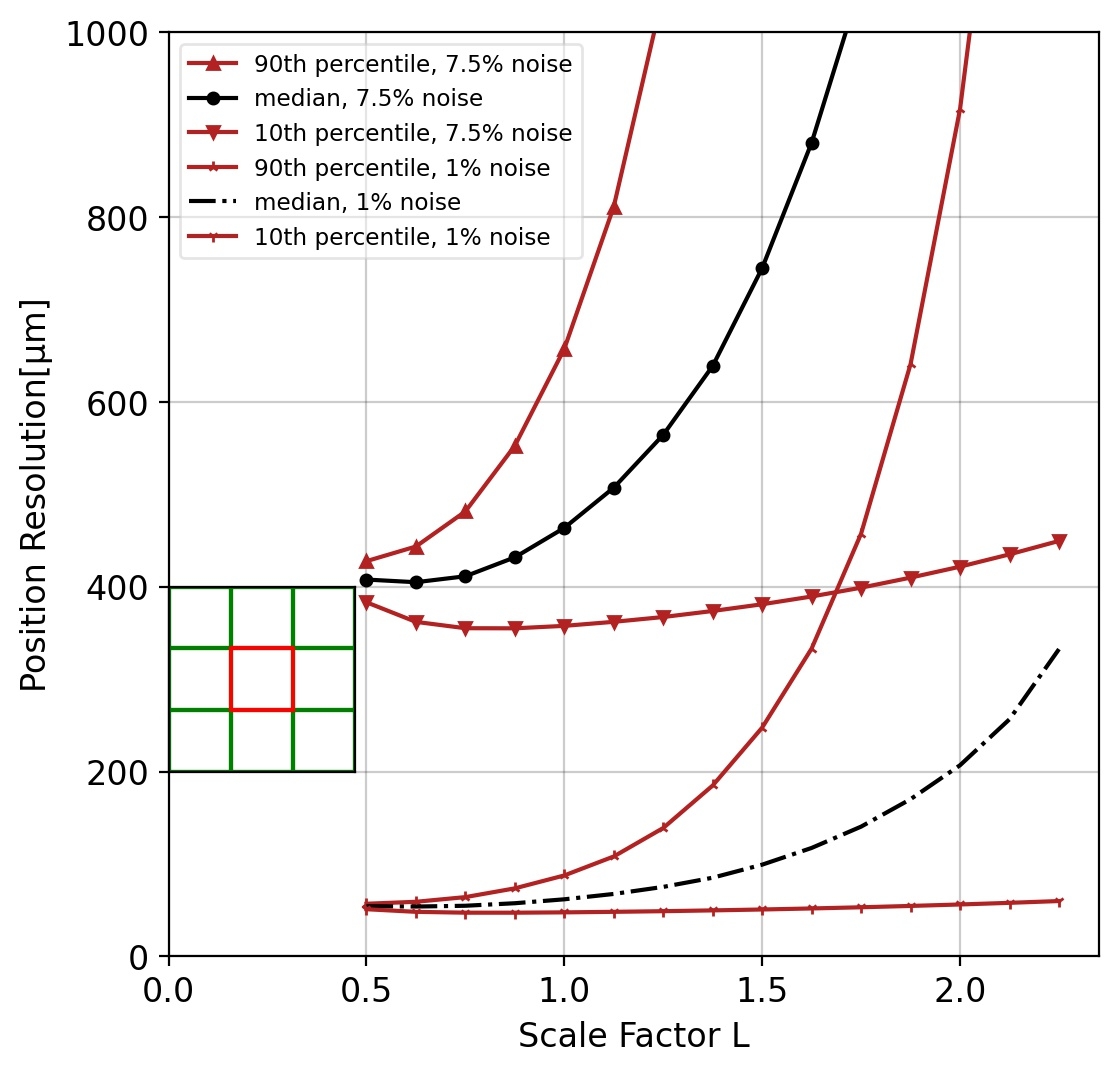}
\includegraphics[width=0.4\textwidth]{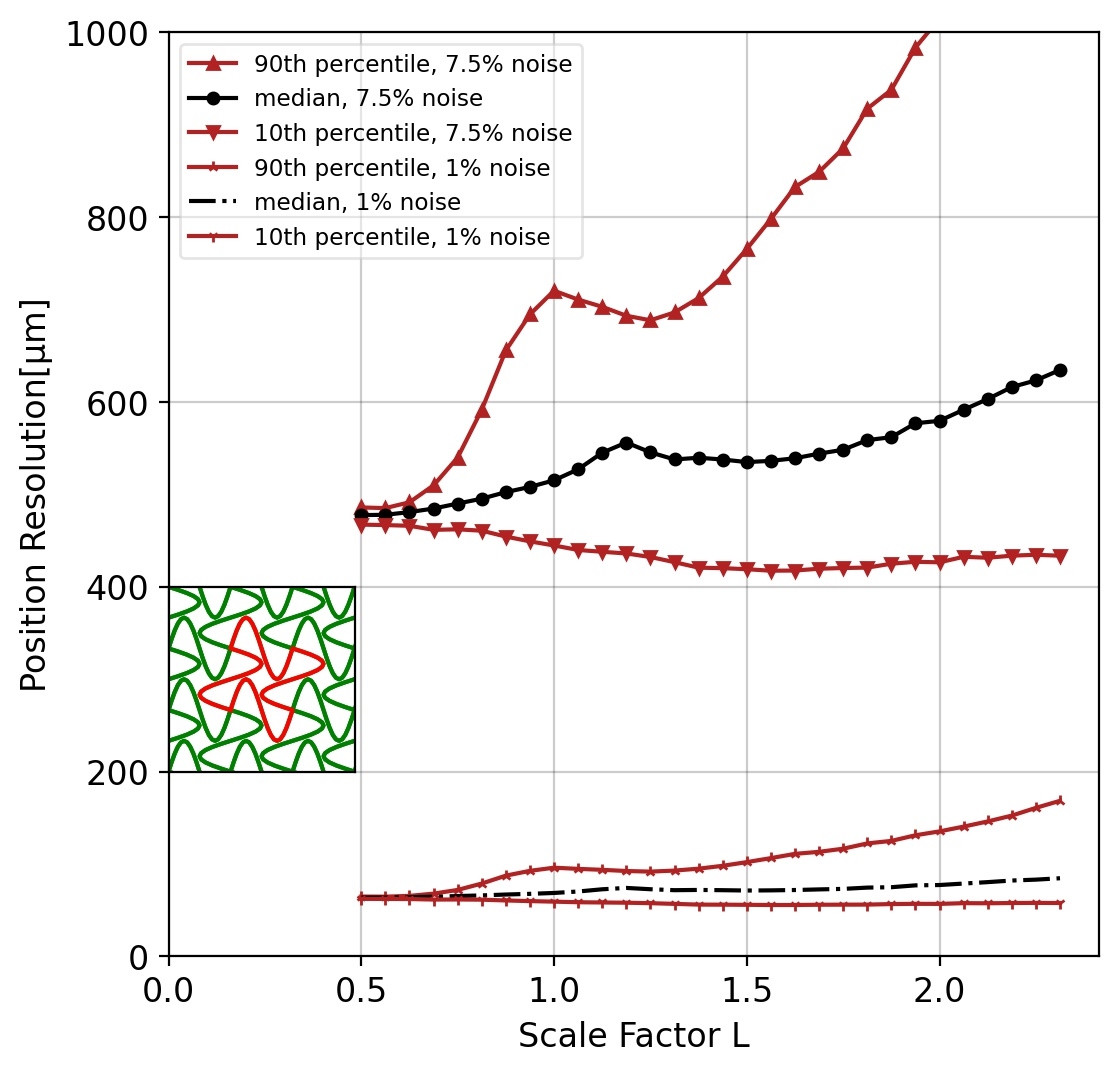}
\caption{The median, 10th, and 90th percentile of the 2D position resolution as a function of the scale factor $L$. The top set of curves have a noise level of 7.5\% and the bottom set of curves have a noise level of 1\%. Left: Square pad pattern. Right: Sinusoidal pad pattern.} \label{fig:square_sine_resolution_vs_scale}
\end{figure}

The scale factor $L$ determines the number of pads and hence the channel count per unit area. A scale factor of $L=1.0$ corresponds to no net gain over a simple square pattern in channel count. A scale factor of $L>1$ corresponds to a reduction in channel count.  Figure~\ref{fig:square_sine_resolution_vs_scale} shows the median, 10th, and 90th percentile of the 2D position resolution function as $L$ is varied for square pads (Left) and sinusoidal pads (Right).

While the two pad patterns have comparable resolutions at $L$ close to 1, the sinusoidal pattern outperforms the square pattern out to $L = 2$. The divergence of the resolution in the square pattern at large $L$ comes from the image being fully contained within single pads with no sharing to constrain the position.

\FloatBarrier
\section{Distributed Pads Using Pickup Internal Layers}

There are applications that require large-area photo-coverage but have low occupancies~\cite{occupancy_vs_rate}, and for which time resolutions less than 100 psec are adequate~\cite{Andrey_Boron8_2020}. These applications are natural for RF strip-line readouts~\cite{Tang_Naxos,timing_paper,Angelico_thesis}, which are 1-dimensional and for which the channel count scales linearly with area rather than quadratically.

However, the signal pickup board can be economically and quickly implemented as a multi-layer printed-circuit (PC) card, allowing multiple internal signal and ground layers to connect physically non-adjacent pads to produce a single electrical pad through vias and internal traces. These distributed pads can connect through vias and traces to front-end digitization electronics directly on the back of the signal card. Values of the scale factor $L$ substantially less than 1.0 can also be explored.

The advantages for high-rate low-occupancy applications will depend on the individual application. Disadvantages include RF impedance mismatches, higher capacitance, and higher cross-talk. However the distributed pad solution may reduce the channel count for applications with a small charge image or requiring custom pad shapes.

\begin{figure}[!ht]
\centering
\includegraphics[width=0.50\textwidth]{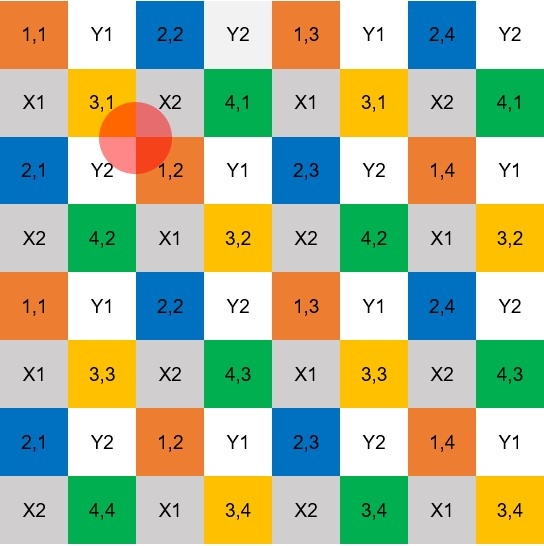}
\caption{An implementation of distributed pads on the pickup board. Half of the cells each have 2 indices, with the first index also represented by the color of the cell. Cells with the same two indices are connected together using internal layers on the printed circuit board. The other half are each connected to one of the four channels, X1, X2, Y1, Y2. In total, there are 20 distinct channels in this pattern. The size of the anode charge pattern is indicated by the disk in the upper left quadrant, corresponding to a scale factor of $L=1.0$.}
 \label{Fig:distributed pads}
\end{figure}

Figure~\ref{Fig:distributed pads} shows an example implementation (the `Park' grid). Half of the cells each have 2 indices, with the first index also represented by the color of the cell. Cells with the same two indices are connected together using internal layers on the printed circuit board. The other half are each connected to one of the four channels, X1, X2, Y1, Y2. There are 64 pads total in this 8-by-8 array of cells. The first indices repeat with a period of 4 pads both vertically and horizontally. In odd-numbered columns, the second index remains constant across the whole column. In even-numbered rows, the second index remains the same. The channels X1, X2, Y1, and Y2 appear with a period of 4 pads both vertically and horizontally. This pattern has a different scaling of channel count to digitized area, with an $8\times8$ array corresponding to 20 channels, a $16\times16$ array corresponding to 36 channels, and a $32\times32$ array corresponding to 68 channels. In general, a $4n\times4n$ array has $8n+4$ channels.


\section{Summary}
\label{summary}

The development of large-area MCP-PMT photodetectors has opened the possibility of applications with photocoverage measured in tens or hundreds of square meters with sub-millimeter spatial resolutions and time resolutions measured in tens of picoseconds. For high-rate applications, such as medical imaging and high-energy particle colliders, a highly-segmented readout is required. Thus an anode geometry consisting of pads is preferred over a strip geometry with a lower channel count. Incorporating a capacitively-coupled anode in the MCP-PMT allows complex patterns of pads to be easily implemented on a
printed circuit card external to the vacuum package.

In a geometry consisting of an array of pads, the number of electronics channels is proportional to the area covered. We present here an example of the use of charge sharing among pads to lower the channel count per unit area. The results from simulations are presented with the pad size scaled to the charge image at the anode from Cherenkov radiation from a charged particle or by a single photon. The simulated signal is represented as a signal image with intensity that varies as a Gaussian and centered on the particle position. The signal sharing is calculated using the magnitude of overlap of this image with all pads in an anode pattern.

The area in which sharing occurs can be enhanced with patterns having convex/concave boundaries. A conventional pattern of regular square pads serves as the baseline for comparison of spatial resolution and channel count per unit area. The patterns are scaled to the diameter of the charge image, with a scaling factor $L$ defined as the ratio of the horizontal or vertical extent of the pad to the 4$\sigma$ diameter of the Gaussian signal image. A value of $L$ greater than 1.0 indicates a larger pad, and hence a lower channel count per unit area.

Noise from digitizing electronics, taken here as 7.5\% and 1\% of the total signal induced by the charge cloud, increases the amount of overlap with neighboring pads required for a given spatial resolution.

As an example we present the simulation of a pattern with pad boundaries formed by horizontal and vertical sine functions. At a scaling factor of $L=1.0$ the pattern returns a maximum spatial resolution for incident single photons of $\approx 800$  microns over the full area, similar to that of the baseline square pad pattern.  However, the sine pattern performs better as $L$ increases. The pattern performs at $L=1.5$ still with a maximum of 800 microns,  while the resolution of the square pattern diverges past 1000 microns for $L > 1.5$. The sine pattern allows a reduction in channel count by a factor of 2.25 with a typical resolution of $\approx600$ microns.

Capacitive coupling of the monolithic internal anode to the pattern of electrodes on an external signal pickup board allows the use of inexpensive, widely-available,  and fast turn-around printed-circuit technology.
For low occupancy applications, multi-layer printed circuit boards allow connecting non-adjacent pads in patterns that uniquely encode the position of the charge pattern. The encoded multi-layer pickup has the property that the number of channels scales linearly in the number of pads per linear length rather than quadratically as in the adjacent-pad case.

\section{Acknowledgements}
\label{acknowledgements}

This work was supported by the Nuclear Physics Division of the Department of Energy through award number DE-SC0015267 and  by the High Energy Division through awards DE-SC-0008172 and DE-SC-0020078. E. Angelico gratefully acknowledges funding by the DOE Office of Graduate Student Research (SCGSR) program, managed by ORAU under contract number DE-SC0014664. All opinions expressed in this paper are the author's and do not necessarily reflect the policies and view of DOE, ORAU, or ORISE. J. Park and F. Wu thank the Physical Sciences Division, Enrico Fermi Institute (EFI),  and the College of the University of Chicago.  We thank Mary Heintz of the EFI for superlative computing and electronics support.

\section*{Appendix A: Calculating the Signal Distribution Gradient}
\label{Mathematics}

A discrete look-up table of signal distributions is used for calculating the signal-spatial gradient at each location on the anode pattern.
We denote $(x,y)$ as the position of the center of the charge cloud, or particle position, induced by the Cherenkov photons. The fraction of total signal that is measured on the $i$th pad is denoted as $P_i$, and is calculated based on the shape of the pad and the particle position:
\begin{equation}
P_i = P_i(x,y)
\end{equation}\\
The signals induced by particles with positions $x_j$ and $y_k$ are numerically simulated and stored as a look-up table. The entries in the look-up table that represent the charge collected by pad $i$ at each position $(x_j,y_k)$ are defined as $T_i(x_j, y_k)$.

In a real (i.e. not simulated) detection event, the charge shower may land at location $(x,y)$ which is close to but not exactly at a simulated point in the look-up table, $(x_j,y_k)$. If the two points are close enough, the following holds by linear approximation:\begin{equation}
	P_i(x,y) = T_i(x_j,y_k) + \pdv{x}T_i(x_j, y_k)(x - x_j) + \pdv{y}T_i(x_j, y_k)(y-y_k)
\end{equation}
Considering the equation above for all pads, we define the following matrix form:
\begin{equation}
	\begin{bmatrix}
	\dots\\
	P_i(x,y)- T_i(x_j,y_k)\\
	\dots
	\end{bmatrix}=\begin{bmatrix}
	\dots&\dots\\
	\pdv{x}T_i(x_j, y_k) & \pdv{y}T_i(x_j, y_k)\\
	\dots&\dots
	\end{bmatrix}
	\begin{bmatrix}
	x-x_j\\
	y-y_k
	\end{bmatrix}
\end{equation}\begin{equation}
	\Delta\mathbf{P} = \mathbf{J}\Delta\vec{r}
\end{equation} Since we want to calculate $\Delta\vec r$ from $\Delta\mathbf{P}$, we want a matrix $\mathbf{K}$ such that $$\mathbf{KJ} =\begin{bmatrix}
1&0\\
0&1
\end{bmatrix} $$which implies
\begin{equation}
	\Delta\vec{r} = \mathbf{K}\Delta\mathbf{P}.
\end{equation} \\
Define $$\pd _x \vec{T} = \begin{bmatrix}
\dots\\
\pdv{x}T_i(x_j, y_k)\\
\dots
\end{bmatrix},\pd _y \vec{T} = \begin{bmatrix}
\dots\\
\pdv{y}T_i(x_j, y_k)\\
\dots
\end{bmatrix}$$If $T_i(x_j, y_k)$ is defined as $T_{i, j, k}$ for short, $\pd _x \vec{T}$ and $\pd _y \vec{T}$ at each point could be calculated discretely:

\begin{equation}
	\pdv{x}T_i(x_j, y_k) \simeq \frac{T_{i,j+1, k}-T_{i,j, k}}{x_{j+1}-x_j}
\end{equation}
\begin{equation}
	\pdv{y}T_i(x_j,y_k) \simeq \frac{T_{i,j, k+1}-T_{i,j, k}}{y_{k+1}-y_k}
\end{equation}We rewrite $\mathbf{J}$ using $\pd _x\vec{T}$ and $\pd _y\vec{T}$:$$ \mathbf{J} = \begin{bmatrix}
\pd _x\vec{T} & \pd _y \vec{T}
\end{bmatrix}$$$\mathbf{K}$ may be calculated from the formula for the matrix inverse, only when $\pd _x\vec T$ is not parallel with $\pd _y\vec T$:
\begin{equation}
	\mathbf{K} = \frac{1}{(\pd _x \vec{T})^2(\pd _y \vec{T})^2-(\pd _x \vec{T}\cdot\pd _y \vec{T})^2}\begin{bmatrix}
	(\pd _y \vec{T})^2 & -\pd _x \vec{T}\cdot\pd _y \vec{T}\\
	-\pd _x \vec{T}\cdot\pd _y \vec{T} & (\pd _x \vec{T})^2\\
	\end{bmatrix}\mathbf{J}^T
\end{equation} \\

\subsection*{A.1: Discrete computation of $(x,y)$ and $(\sigma_x, \sigma_y)$}
\label{Computation}
Given a set of signals $\vec{P}$ collected by the pads in a measured charged-particle event, called the ``signal distribution'', one may find the particle position in a simulated look-up table that produces the smallest deviation from the observed signal distribution. If the signal collected by the $i$th pad from a particle impinging on position $(x_j, y_k)$ is denoted in the look-up table $\vec{T}$ as $T_i(x_j, y_k)$, then the reconstructed position of the particle $\vec{r}$ is estimated by interpolating the deviation from the best-fit location $(x_j, y_k)$ using the inverse-gradient matrix, $\mathbf{K}$, from Appendix \ref{Mathematics}:
\begin{equation}
	\vec{r} =\begin{bmatrix}
	x_j\\y_k
	\end{bmatrix} + \mathbf{K}(x_j,y_k)(\vec{P}(x, y) - \vec{T}(x_j, y_k))
\end{equation}\\
The matrix $\mathbf K$ depends on the geometry of the pad pattern used. It diverges at positions where $\pd _x\vec T$ is parallel with $\pd _y\vec T$, i.e. when a variation of the photon positions causes no change in the measured signal on the pads. In this case, the position resolution at that point is considered to be infinite and the pattern is dubbed `degenerate'.

When the anode pattern is not degenerate, the uncertainty of the measurement of the signal distribution, $\vec P$, results in an uncertainty in the reconstructed position, $\vec r$. Even if the look-up table were infinitesimally discretized and the shape of the charge-shower were perfectly known, $\vec{r}$ would have some uncertainty cause by voltage noise of the digitizing electronics. The noise is characterized by a fractional RMS of $\sigma_C$ at the percent level. The position uncertainties in $x$ and $y$ on $\vec{r}$, $\sigma_x$ and $\sigma_y$, are then related to the charge noise by

\begin{equation}
\sigma^2_x = \sum_{i = 1}^{\text{\# of pads}}(K_{0i}(x_j, y_k))^2\sigma^2_{C}
\end{equation}
\begin{equation}
\sigma^2_y = \sum_{i = 1}^{\text{\# of pads}}(K_{1i}(x_j, y_k))^2\sigma^2_{C}
\end{equation}\\

The quantities in the main text that refer to position resolution are $\sigma = \sqrt{\sigma_x^2 + \sigma_y^2}$. The noise $\sigma_C$ may be factored out so that the resolution functions reported may be scaled to be useful for other experimental setups with other noise factors.



\begin{thebibliography}{99}
\bibitem{units_c_equal_1}
We use `natural' units in which the speed-of-light $c$ is defined to $c=1.0$, so space is measured in the same units as time. For example, 10 psec is $~3$ mm of light travel time and vice versa
\bibitem{occupancy_vs_rate}
We define occupancy as the number of hits per channel per resolving time. The limitation on occupancy is an orthogonal constraint to the limitation in rate, and depends on the multiplicity in the signal source (e.g. a high-energy hadron shower of thousands of simultaneous particles at one extreme and a single optical photon at the other), the pad size, and the resolving time, For low-occupancy high-rate applications a distributed pad construction can extend performance into the MegaHz/channel region with $\le 10^{-3}$ loss due to multiple hits in a channel.
\end{thebibliography}
\end{document}